\documentclass[onecolumn,showpacs,nobibnotes,nofootinbib, greek, hebrew]{revtex4}

\usepackage{amsmath,amssymb}
\usepackage{graphicx}
\usepackage{subfigure}

\newcommand{\be}{\begin{equation}}
\newcommand{\ee}{\end{equation}}
\newcommand{\bea}{\begin{eqnarray}}
\newcommand{\eea}{\end{eqnarray}}

\def\beq{\begin{equation}}
\def\eeq{\end{equation}}

\def\beqn{\begin{eqnarray}}
\def\eeqn{\end{eqnarray}}

\begin{document}

\title{ On  the Analytic Solution of  the  Balitsky-Kovchegov Evolution Equation }
\author{Sergey  Bondarenko and Alex Prygarin  }
%\email{prygarin@gmail.com}
\affiliation{ \vspace{0.5cm} Department of Physics, Ariel University, Ariel 40700, Israel}
\pacs{ 12.38.Bx,  12.40.Nn, 11.55.Jy,  24.85.+p}

\begin{abstract}
 The study presents an analytic solution of the Balitsky-Kovchegov~(BK) equation in a particular kinematics. The solution is written in  the momentum space and based on the eigenfunctions of the  truncated   Balitsky-Fadin-Kuraev-Lipatov~(BFKL) equation  in the gauge adjoint representation, which was used for calculation of the Regge~(Mandelstam) cut contribution to the planar helicity amplitudes. We introduce an eigenfunction of the singlet BFKL equation constructed of  the adjoint eigenfunction multiplied  by a factor, which restores the dual conformal symmetry present in the  adjoint and broken in the singlet BFKL equations.
  The proposed analytic BK solution correctly reproduces the initial condition and the high energy asymptotics of the scattering amplitude.

\end{abstract}

\maketitle

\section{Introduction}\label{sec:intro}

The power growth of the Balitsky-Fadin-Kuraev-Lipatov~(BFKL)~\cite{bfkl} amplitude contradicts the unitarity condition and should be modified. In the pioneering paper of Gribov, Levin and Ryskin~(GLR)~\cite{glr} the authors introduced a gluon saturation,  meaning that  the gluon density grows at large energy resulting in an overlap of the gluon states  which tams  growth of the amplitude with increase of energy. This mechanism introduces a new scale commonly referred to as a saturation scale $Q^2_s$. The  GLR equation was first to account for non-linear effects for describing  the physics   of the Deep Inelastic Scattering~(DIS). The GLR equation has a linear term representing BFKL type ladder diagrams, but also it includes a non-linear term, which stands for so-called fan diagrams describing the pomeron splitting into two pomerons. The GLR equation was formulated in double logarithmic kinematics, where one resums powers of logarithm of energy and typical transverse momentum. This kinematics is more restricted with respect to the BFKL multi-Regge kinematics, where the energy of the colliding particles is much larger than any transverse momentum.
A decade after formulation of the GLR equation the  unitarization corrections in the Regge kinematics were taken into account in the further generalization of the BFKL equation in the Wilson line formalism by Balitsky~\cite{bal} and soon after that independently by Kovchegov~\cite{kov} in the color dipole approach~\cite{dipole, mueller}. The Balitsky chain reduced to the equation derived by Kovchegov  in the limit of infinite number of colors. The resulting Balitsky-Kovchegov~(BK) equation has a relatively simple form in the transverse coordinate space where the integral Kernel of the linear and the non-linear terms are happened to be the same and  have a simple  interpretation of splitting of the color dipole into two dipoles. Despite numerous attempts during almost two decades the full analytic solution to the BK equation is still to be found.
  The linear part of the BK equation corresponds to the BFKL evolution, which was completely solved exploiting conformal symmetry of the BFKL Kernel in the space of transverse coordinates. However,  the direct substitution of the BFKL eigenfunctions in the coordinate space to the BK equation does not lead to significant  simplifications in solving the BK equation.
It was recently proposed that BFKL equation in color adjoint  representation  can be solved for non-zero transverse momentum using the eigenfunction in the momentum space. This was done for calculating the Regge~(Mandelstam) cut contribution to the planar helicity amplitudes in Regge kinematics. A resulting series of publications confirmed this calculations to be correct by performing analytic continuation followed by Regge limit of the so-called remainder function of the Bern-Dixon-Smirnov~(BDS)~\cite{bds} scattering amplitude. The analytic solution of the  color adjoint BFKL was also useful for calculating Regge cut corrections to the planar amplitudes of the  next-to-maximally helicity violating~(NMHV) configuration of the  scattered particles~\cite{prygarinnmhv}. The Moebius representation of the adjoint BFKL equation was calculated in ref.~\cite{moebiusAdjoint}.

In the present study we build a general form of the analytic solution to the BK equation~(see eq.~(\ref{BKsol})) based on the eigenfunctions of the color adjoint BFKL equation. The difference between the eigenfunctions of the  adjoint BFKL  in eq.~(\ref{fkq}) and the eigenfunctions of the singlet BFKL  in eq.~(\ref{Fkq}) is  mainly in the proper normalization condition leading to a factor, which is related to a  shift the argument of the BFKL eigenvalue by $1/2$ with respect the eigenvalue of  adjoint BFKL. This factor is needed to restore the conformal symmetry in the momentum space of dual coordinates $p_i=z_i-z_{i+1}, \; i=1,2$ present in the adjoint BFKL and broken in the singlet case. The dual conformal symmetry is believed to be a sign of integrability in planar amplitudes and is supposed to be present in the integrable singlet BFKL equation.

  Next, assuming a particular dependence of the impact factor on the transferred momentum in eq.~(\ref{PhiNotq}) we find a closed analytic solution of the Balitsky-Kovchegov equation in this particular kinematics.  Finally, we check that our solution correctly reproduces the initial condition and  the high energy asymptotics of the scattering amplitude.
Some details of the calculations are presented in the Appendix.

\section{Solution  of the BK equation}\label{app:lin}

The Balitsky-Kovchegov~(BK)~\cite{bal, kov} equation describes the energy evolution of  the  imaginary part of the scattering amplitude $N(\mathbf{x},\mathbf{y})$ as follows
\beqn\label{BKx}
 \partial_Y N(\mathbf{x},\mathbf{y}) = \frac{\bar{\alpha}_s}{ 2 \pi} \int \frac{d^2 \mathbf{z} \; (\mathbf{x}-\mathbf{y})^2}{(\mathbf{x}-\mathbf{z})^2 \; (\mathbf{z}-\mathbf{y})^2}
\left( \frac{}{}
 N(\mathbf{x},\mathbf{z})+  N(\mathbf{z},\mathbf{y})- N(\mathbf{x},\mathbf{y}) -N(\mathbf{x},\mathbf{z})N(\mathbf{z},\mathbf{y})
 \frac{}{}\right).
\eeqn
The BK equation is written in terms of `t Hooft coupling
\beqn
\bar{\alpha}_s=\frac{\alpha_s N_c}{\pi} % 9.170 LipatovBook
\eeqn
and the rapidity
\beqn
Y=\ln \left(\frac{s}{s_0}\right),
\eeqn
where $s$ is the square of the total energy of the scattering particles  and $s_0$ is  some energy scale.
Numerous attempts of solving BK equation in coordinate space    suggest us looking for analytic solution in other representations. It is natural to start with momentum space using a notation of ref.~\cite{marquet} for the corresponding Fourier transform
\beqn\label{Fourier}
\mathcal{N}(\mathbf{k},\mathbf{q})=  \int \frac{d^2 \mathbf{x}}{2\pi} \; \frac{d^2 \mathbf{y}}{2 \pi} \; e^{i \mathbf{k} \mathbf{x} } e^{i (\mathbf{q}-\mathbf{k}) \mathbf{y}} \frac{N (\mathbf{x}, \mathbf{y})}{(\mathbf{x}-\mathbf{y})^2}
\eeqn
and write the Balitsky-Kovchegov equation in the momentum space as follows
\beqn\label{BKmom}
\partial_Y \mathcal{N} (\mathbf{k}, \mathbf{q})=\frac{\bar{\alpha}_s}{\pi} \int \frac{d^2 \mathbf{k}'}{(\mathbf{k}-\mathbf{k}')^2}
\left(
\mathcal{N}(\mathbf{k}', \mathbf{q})
-\frac{1}{4} \left[ \frac{(\mathbf{q}-\mathbf{k})^2}{(\mathbf{q}-\mathbf{k}')^2}  + \frac{\mathbf{k}^2}{{\mathbf{k}'}^2}\right]\mathcal{N}(\mathbf{k}, \mathbf{q})
\right)
-\frac{\bar{\alpha}_s}{2 \pi} \int d^2 \mathbf{k}' \mathcal{N} (\mathbf{k}, \mathbf{k}')
 \mathcal{N} (\mathbf{k}-\mathbf{k}', \mathbf{q}-\mathbf{k}').
\eeqn

Before considering the non-linear evolution we go back to the linear BFKL equation, which can be solved in two different ways; in the coordinate and in the momentum spaces. Both solutions are well known and usually addressed to as $q=0$ solution (momentum  space) and $q\neq 0$ solution~(coordinate space). Those two were thoroughly discussed in numerous studies and we  do not focus on them for the sake of brevity. For more details on these two solutions the reader is referred to  review texts~\cite{texts}.
This is true for the  BFKL equation projected on the color singlet state of two interacting reggeized gluons. However, if one considers color adjoint BFKL with some infra-red divergent parts removed, there is  another $q\neq 0$ solution in the momentum space expressed through the  adjoint  BFKL eigenfunctions\footnote{Here we use complex coordinates defined by  $k=k_x+i k_y, \;   k^*=k_x-i k_y$}
\beqn\label{fkq}
f_{\nu,n} (k,q)=\left(\frac{k}{q-k}\right)^{i \nu +\frac{n}{2}}\left(\frac{k^*}{q^*-k^*}\right)^{i \nu -\frac{n}{2}},
\eeqn
which were used in a serious of publications~\cite{remBDSregge} for calculations of the Regge~(Mandelstam) cut contribution to the helicity amplitudes in maximally supersymmetric theory. Its solution was written in terms  of $f_{\nu,n} (k,q)$ and a corresponding eigenvalue of the adjoint BFKL equation of shifted argument with respect to the singlet BFKL eigenvalue.
 The corresponding adjoint BFKL Green function was investigated in ref.~\cite{Chachamis:2012fk}.

 In the present study we consider  a  non-linear generalization of the linear BFKL equation, namely the Balitsky-Kovchegov~(BK) evolution equation. We introduce an ansatz for the BK solution in a particular kinematic regime, which can be written in terms of functions similar to $f_{\nu,n}(k,q)$. We slightly redefine the   "adjoint"  $f_{\nu,n}(k,q)$   to account for   a proper normalization of the  singlet BFKL as well as  a minus sign, which happens to be important for the non-linear  term of the BK equation as follows
\beqn\label{Fkq}
F_{\nu,n} (k,q)=\frac{1}{ \sqrt{2} \; \pi} \frac{1}{|k||k-q|}
\left(\frac{k}{k-q}\right)^{i \nu +\frac{n}{2}}
 \left(\frac{k^*}{k^*-q^*}\right)^{i \nu -\frac{n}{2}}
 =\frac{1}{ \sqrt{2} \; \pi} \frac{(-1)^n}{|k||k-q|} f_{\nu,n} (k,q).
\eeqn
The function in eq.~(\ref{Fkq}) solves the leading order singlet BFKL equation as we show below.
The major difference between the adjoint and singlet eigenfunctions is   the factor $\frac{1}{|k||k-q|}$, which restores the dual conformal symmetry of the singlet BFKL equation. This analysis is beyond the scope of the present study and will be published by us elsewhere.

The corresponding  orthogonality  condition is given by

\beqn
\int d^2 \mathbf{k} \;  F^*_{\nu,n} (k, q) \; F_{\nu',n'} (k, q)=  \frac{\delta(\nu -\nu')\delta_{n, n'}}{|q|^2}
\eeqn
and the completeness condition
reads
\beqn\label{comp}
\sum_{n=-\infty}^{\infty} \int_{-\infty}^{\infty} d \nu \; F^*_{\nu, n} (k', q) \; F_{\nu, n} (k, q)=\frac{\delta^2 (k-k')}{|q|^2}.
\eeqn
Then we can write  the  analytic solution of the Balitsky-Kovchegov equation in the momentum space as follows
\beqn\label{BKsol}
\mathcal{N} (\mathbf{k},\mathbf{q}) = \sum_{n=-\infty}^{\infty} \int_{-\infty}^{\infty}  d \nu \;    C_{\nu,n}(Y) \;F_{\nu ,n }(k,q),
\eeqn
where the coefficient function $C_{\nu,n}(Y)$ is found  below plugging eq.~(\ref{BKsol}) in the BK equation. Here we make a basic assumption that the coefficient function $C_{\nu,n}(Y)$ does not depend on $k$ or $q$. We discuss this approximation in more details below, where we consider the initial condition and the high energy limit of the obtained BK solution.

It is useful to introduce new complex variables
\beqn
\frac{k}{k-q}=w, \; \frac{k^*}{k^*-q^*}=w^*.
\eeqn
In terms of  the  coordinates  $w$ and $w^*$  the function in eq.~(\ref{Fkq}) reads
\beqn\label{Fw}
F_{\nu,n} (k,q)=\tilde{F}_{\nu,n} (w,q)=\frac{1}{ \sqrt{2} \; \pi}\frac{|w-1|^2}{|q|^2} w^{i \nu +\frac{n}{2}-\frac{1}{2}}
{w^*}^{i \nu -\frac{n}{2}-\frac{1}{2}}
\eeqn

and  the completeness relation in eq.~(\ref{comp}) is given by
\beqn\label{compw}
&&
\sum_{n=-\infty}^{\infty} \int_{-\infty}^{\infty} d \nu \; \tilde{F}^*_{\nu, n} (w', q) \; \tilde{F}_{\nu, n} (w, q)=
\frac{|w-1|^4}{|q|^4}\delta^2 (w- w').
\eeqn

Consider the first  linear term of the BK equation in the momentum space eq.~(\ref{BKmom})

\beqn\label{Fkqw}
\frac{\bar{\alpha}_s}{2 \pi}
\int \frac{d^2 \mathbf{k}'}{(\mathbf{k}-\mathbf{k}')^2} \mathcal{N} (\mathbf{k}', \mathbf{q})=\frac{\bar{\alpha}}{2 \pi}
\sum_{n=-\infty}^{\infty} \int_{-\infty}^{\infty} d\nu  \;  C_{\nu,n} (Y)\; \int \frac{d^2 \mathbf{k}'}{(\mathbf{k}-\mathbf{k}')^2}\; F_{\nu,n}(k',q)
\eeqn
and calculate the following expression
\beqn
&& \int \frac{d^2 \mathbf{k}'}{(\mathbf{k}-\mathbf{k}')^2} F_{\nu,n}(k',q)=
\int \frac{d^2 \mathbf{k}'}{(\mathbf{k}-\mathbf{k}')^2}
\frac{1}{\sqrt{2} \;\pi}\frac{1}{|k'||k'-q|}\left(\frac{k'}{k'-q}\right)^{i \nu +\frac{n}{2}}\left(\frac{{k'}^*}{{k'}^*-q^*}\right)^{i \nu -\frac{n}{2}}
=
\\
&&  \frac{1}{ \sqrt{2} \; \pi}\frac{|w-1|^2}{|q|^2} \int \frac{d^2 \mathbf{w'}}{|w-w'|^2} \frac{{w'}^{i \nu +\frac{n}{2}}  {{w'}^*}^{i \nu -\frac{n}{2}}}{|w'|} \nonumber
=
\frac{1}{\sqrt{2} \;\pi}\frac{|w-1|^2}{|q|^2}\frac{w^{i \nu +\frac{n}{2}}{w^*}^{i \nu -\frac{n}{2}}}{|w|}
 \int \frac{d^2 \mathbf{z} \;  z^{i\nu +\frac{n}{2}-\frac{1}{2}}
{z^*}^{i\nu -\frac{n}{2}-\frac{1}{2}}}{|1-z|^2}
 =
\nonumber
\\
&& F_{\nu,n}( k, q) \int \frac{d^2 \mathbf{z}}{ |1-z|^2}
z^{i\nu +\frac{n}{2}-\frac{1}{2}}
{z^*}^{i\nu -\frac{n}{2}-\frac{1}{2}} \nonumber
\eeqn
with  $z=w'/w$. Thus we have

\beqn
\frac{\bar{\alpha}_s}{\pi}\int \frac{d^2 \mathbf{k}'}{(\mathbf{k}-\mathbf{k}')^2} \mathcal{N} (\mathbf{k}', \mathbf{q})=\frac{\bar{\alpha}_s}{\pi}
\sum_{n=-\infty}^{\infty} \int_{-\infty}^{\infty} d\nu \;  C_{\nu,n} (Y)\;
F_{\nu,n}( k, q) \int \frac{d^2 \mathbf{z}}{ |1-z|^2} \;
z^{i\nu +\frac{n}{2}-\frac{1}{2}}
{z^*}^{i\nu -\frac{n}{2}-\frac{1}{2}}.
\eeqn

In a similar way  the second linear term of the BK equation  in eq.~(\ref{BKsol}) can be written as
\beqn
-\frac{1}{4}\frac{\bar{\alpha}_s}{\pi}\int \frac{d^2 \mathbf{k}'}{(\mathbf{k}-\mathbf{k}')^2 } \frac{(\mathbf{q}-\mathbf{k})^2}{(\mathbf{q}-\mathbf{k}')^2}
\mathcal{N}(\mathbf{k}, \mathbf{q} )=
 -\frac{1}{4}\frac{\bar{\alpha}_s}{\pi}
 \sum_{n=-\infty}^{\infty} \int_{-\infty}^{\infty} d\nu \; C_{\nu,n}(Y) \; F_{\nu,n}( k, q)  \int \frac{d^2 \mathbf{z}}{|1-z|^2}
  \eeqn
and finally the third linear term  reads

\beqn
-\frac{1}{4}\frac{\bar{\alpha}_s}{\pi}\int \frac{d^2 \mathbf{k}'}{(\mathbf{k}-\mathbf{k}')^2 } \frac{ \mathbf{k}^2}{{\mathbf{k}'}^2}
\mathcal{N}(\mathbf{k}, \mathbf{q})=
-\frac{1}{4}\frac{\bar{\alpha}_s}{\pi}
\sum_{n=-\infty}^{\infty} \int_{-\infty}^{\infty} d\nu \; C_{\nu,n}(Y) \; F_{\nu,n}( k, q)  \int \frac{d^2 \mathbf{z}}{|z|^2|1-z|^2}.
\eeqn
The sum of these three linear terms gives an expression, which determines the BFKL eigenvalue
\beqn\label{omegaBFKL}
\omega(\nu,n)=\bar{\alpha}_s \chi(\nu,n)
\eeqn
through a function
\beqn\label{chi} \hspace{-0.5cm}
\chi(\nu,n)=
\int \frac{d^2 \mathbf{z}}{|1-z|^2} \;
\left(z^{i\nu +\frac{n}{2}+\frac{1}{2}}
{z^*}^{i\nu -\frac{n}{2}-\frac{1}{2}}-\frac{1}{|z|^2+|1-z|^2}\right)
=
2 \psi(1)-\psi\left(i \nu +\frac{|n|}{2}+\frac{1}{2}\right)-\psi\left(-i \nu +\frac{|n|}{2}+\frac{1}{2}\right),
\eeqn
which is
expressed in terms of the digamma function defined as a logarithmic derivative of Euler gamma function
\beqn
\psi(z)=\frac{d \ln \Gamma(z)}{ d z}=\int_{0}^1 dx \frac{x^{z-1}}{x-1}.
\eeqn
In deriving eq.~(\ref{chi}) we used the identity
\beqn
\int \frac{d^2 \mathbf{z}}{|z|^2 |1-z|^2}=2 \int \frac{d^2 \mathbf{z}}{ |1-z|^2  \left(|z|^2+ |1-z|^2\right)}.
\eeqn
For more details on the  calculation of $\chi(\nu,n)$ the reader is referred to review texts~\cite{texts}.
This way we have shown that $F_{\nu,n}(k,q)$ in eq.~(\ref{Fkq}) is indeed the BFKL eigenfunction with the eigenvalue $\omega(\nu,n)$ given in commonly used notation without any possible shift of the argument of digamma functions.

Now we are in position of plugging $F_{\nu,n}(k,q)$ into the BK equation in eq.~(\ref{BKmom}). Namely, consider the non-linear term
\beqn
&& -\frac{\bar{\alpha}_s}{2 \pi}\int d^2 \mathbf{k'} \mathcal{N}(\mathbf{k}, \mathbf{k'})\mathcal{N}( \mathbf{k}-\mathbf{k'}, \mathbf{q}-\mathbf{k'})=
\\
&& \nonumber
-\frac{\bar{\alpha}_s}{2 \pi}
 \int d^2 \mathbf{k'}
\sum_{n_1=-\infty}^{\infty}
\sum_{n_2=-\infty}^{\infty}  \int_{-\infty}^{\infty} d \nu_1 \int_{-\infty}^{\infty} d \nu_2
 C_{\nu_1, n_1}(Y)  C_{\nu_2, n_2}(Y) F_{\nu_1, n_1}(k,k') F_{\nu_2, n_2}(k-k', q-k')=
 \\
 &&  \nonumber
 -\frac{\bar{\alpha}_s}{2 \pi}
  \sum_{n_1=-\infty}^{\infty}
\sum_{n_2=-\infty}^{\infty}  \int_{-\infty}^{\infty} d \nu_1 \int_{-\infty}^{\infty} d \nu_2
 \frac{k^{i \nu_1+\frac{n_1}{2}}}{ (k-q)^{i \nu_2+\frac{n_2}{2}}}
  \frac{{k^*}^{i \nu_1-\frac{n_1}{2}}}{ (k^*-q^*)^{i \nu_2-\frac{n_2}{2}}}
  \frac{1}{|k| |k-q|}
 C_{\nu_1, n_1}(Y)  C_{\nu_2, n_2}(Y) \; R_{\nu_1, n_1}^{ \nu_2, n_2},
 \\
 &&
\eeqn
where   $R_{\nu_1, n_1}^{ \nu_2, n_2}$ is calculated as follows (here we define $k/k^*= e^{i 2 \phi}$)
\beqn
&& R_{\nu_1, n_1}^{ \nu_2, n_2} = \int \frac{d^2 \mathbf{k'}}{2 \pi^2} (k-k')^{-i \nu_1 - \frac{n_1}{2}+i \nu_2 + \frac{n_2}{2} -1 }
({k}^*-{k'}^*)^{-i \nu_1 + \frac{n_1}{2}+i \nu_2 - \frac{n_2}{2} -1 }=
\\  \nonumber
&&   \int_{0}^{\infty}  \frac{d   (|k'|^2)}{4 \pi^2} \int_{0}^{2\pi} d \phi  \;\; |k'|^{-i2 \nu_1+i2 \nu_2-2} e^{i \phi (n_2-n_1)} =
 \delta(\nu_2-\nu_1) \; \delta_{n_1,n_2}.
\\  \nonumber
%&& \frac{1}{ 2\pi^2}  \int_{-\infty}^{\infty}  d (\ln |k'|) \; e^{i(2 \nu_2-2 \nu_1) \ln |k'|} \int_0^{2\pi}  d\phi \;  e^{i \phi (n_2-n_1)}=
%\\  \nonumber
%&&   2  \;  \delta( 2 \nu_2-2\nu_1) \delta_{n_1,n_2}=    \delta(\nu_2-\nu_1) \; \delta_{n_1,n_2}
\eeqn

Using the last expression we  write

\beqn\label{nonlin}
  -\frac{\bar{\alpha}_s}{ 2 \pi}\int d^2 \mathbf{k}'  \mathcal{N}(k, k')  \mathcal{N}(k-k', q-k') = -\frac{\bar{\alpha}_s}{\sqrt{2} } \sum_{n=-\infty}^{\infty}
  \int_{-\infty}^{\infty} d \nu \;F_{\nu,n} (k,q) (C_{\nu,n}(Y))^2.
\eeqn
Note that the coefficient of the non-linear term typically associated with the triple Pomeron vertex, in this notation, is  a number independent of anomalous dimension   and conformal spin. Therefore  the ($\nu, n$) dependence of the BK solution enters only through eigenfunctions $F_{\nu, n}(k,q)$  and the BFKL eigenvalue $\omega(\nu,n)$ in eq.~(\ref{Fkq}) and  eq.~(\ref{omegaBFKL}) respectively.
%The constant $\sqrt{2}  \; \pi $  reflects the normalization of the eigenfunctions $F_{\nu,n}(k,q)$.

Using the expression for the non-linear term eq.~(\ref{nonlin}) in the BK equation we get an equation for the coefficient function $C_{\nu,n}(Y)$ as follows
\beqn\label{Cab}
\frac{d C_{\nu,n} (Y)}{ d Y}= a \; C_{\nu,n} (Y)+b \; C_{\nu,n} (Y)^2.
\eeqn

Its solution reads
\beqn\label{Csol}
C_{\nu,n}(Y)= \frac{a C_{\nu,n}(0) \; e^{a Y}}{a +b \;  C_{\nu,n}(0) \left(1 -e^{aY} \right) },
\eeqn
where $a=\omega(\nu,n)$ is the BFKL eigenvalue in eq.~(\ref{omegaBFKL}) and $b=- \bar{\alpha_s}/(\sqrt{2}  )$ is the coefficient of the non-linear  term in eq.~(\ref{nonlin})\footnote{Note that this coefficient depends on  the normalization of the eigenfunctions.}.
The initial condition of  the coefficient function $C_{\nu,n}(Y)$ is set by the following expression
\beqn\label{C0}
 C_{\nu,n}(0)=  |q|^2  \int d^2 k'   \;      F^*_{\nu, n} (k', q) \; \Phi(k', q)
=   |q|^4 \int \frac{d^2 w'}{|w'-1|^4}   \;      \tilde{F}^*_{\nu, n} (w', q) \; \tilde{\Phi}(w', q).
\eeqn

The function $\Phi(k, q)$ is some impact factor  determining an initial condition of the non-linear evolution.  The self-consistency of the approach suggests a choice of the impact factor   describing  multiple rescatterings, e.g.  an impact factor of the Glauber-Mueller type~\cite{glaubermueller}~(see also ref.~\cite{bondarenkobound}). Our basic assumption is   that the coefficient function $C_{\nu,n}(0)$ does not depend on the transferred momentum $q$. This means that we choose the impact factor in such a way that
\beqn\label{PhiNotq}
 \tilde{\Phi}(w', q) =\frac{\tilde{{\Phi}}(w' )}{|q|^2}
\eeqn
in accordance with eq.~(\ref{C0}) and eq.~(\ref{Fw}),  where $\tilde{{\Phi}}(w )$   is some function of $w$ and independent of $q$.
This approximation seems to be reasonable for small values of the transferred momentum.

The evolution equation eq.~(\ref{Cab}) and its solution in eq.~(\ref{Csol}) are similar to that of the phenomenological summation of the Pomeron Fan diagrams
in the Reggen Field Theory for zero transverse momentum~\cite{rft} and can be used for summation of  pomeron loops~\cite{pomeronloops}.
We recall that we wrote a general form of the  BK solution in eq.~(\ref{BKsol}) as
\beqn\label{BKsol2}
\mathcal{N} (\mathbf{k},\mathbf{q}) = \sum_{n=-\infty}^{\infty} \int_{-\infty}^{\infty}  d \nu \;    C_{\nu,n}(Y) \;F_{\nu ,n }(k,q),
\eeqn
 and it  follows directly from this  definition and the completeness condition in eq.~(\ref{comp})  that at $Y=0$ for our choice of $C_{\nu, n} (Y)$ we have
 \beqn
 \mathcal{N} (\mathbf{k},\mathbf{q})|_{Y=0}=  \Phi(\mathbf{k}, \mathbf{q}).
 \eeqn

A general form of the solution $\mathcal{N}(\mathbf{k},\mathbf{q})$ in eq.~(\ref{BKsol2}) together with the expression for the coefficient function
$C_{\nu,n}(Y)$ in eq.~(\ref{Csol}) and its initial condition $C_{\nu,n}(0)$ in eq.~(\ref{C0}) present the main result of the this study.

It is easy to see from eq.~(\ref{BKsol}) and eq.~(\ref{Csol}) that the solution of the BFKL equation in the color singlet state is obtained by  setting $b=0$ in the coefficient function $C_{\nu,n}(Y)$. Note that in the BFKL equation, which corresponds to the linear part of eq.~(\ref{BKmom}) the transferred momentum $q$ is not mixed with the integration variable of the Kernel and thus the condition that $C_{\nu,n}(Y)$ does not depend on $q$ can be relaxed reproducing the full analytic solution of  the BFKL equation for a non-zero transferred momentum.

We have shown  above that our choice of functions $F_{\nu,n} (k,q)$ is consistent with a solution of the BFKL equation, and the next step is to check that the full solution written in terms of these functions  have a reasonable behaviour at high energy.

In the limit $Y \to \infty$ eq.~(\ref{Csol}) reduced to
\beqn
C_{\nu, n} \simeq -\frac{a}{b}=\sqrt{2}  \; \chi(\nu,n)
\eeqn
and the amplitude $\mathcal{N} (\mathbf{k},\mathbf{q})$ defined by eq.~(\ref{BKsol}) in this limit does not depend of the initial condition resulting in
\beqn
\mathcal{N} (\mathbf{k},\mathbf{q})|_{Y \to \infty}  \simeq \sum_{n=-\infty}^{\infty} \int_{-\infty}^{\infty}  d \nu \;  \sqrt{2}  \; \chi(\nu,n)    \;F_{\nu ,n }(k,q)=  \frac{2}{ |q|^2}.
\eeqn
Details of this calculation are presented in the Appendix.
This function peaks at small $q$ and thus can be reasonably approximated by   Dirac $\delta$-function.

 \beqn\label{NYmy}
 \mathcal{N} (\mathbf{k},\mathbf{q})|_{Y \to \infty}\simeq  2  \delta(|q|^2) \delta(\phi)=
  \frac{1}{|q|} \delta(|q|) \delta(\phi)=
 \delta^2 (\mathbf{q}).
\eeqn

Note that we have no $k$ dependence at $Y \to \infty $ and the asymptotic amplitude diverges for  $q=0$. This amplitude represents so called fixed point at which the derivative $\partial_Y  \mathcal{N} (k,q) $ vanishes and the amplitude  goes to its maximal value.
On the other hand one can search fixed points of the   BK equation in  the coordinate space in eq.~(\ref{BKx}) and it is widely accepted that a natural fixed point of eq.~(\ref{BKx}) is $N(\mathbf{x}, \mathbf{y})=1$, which is also a fixed point in phenomenological models. Then for our definition of the Fourier transform in eq.~(\ref{Fourier}) we have

\beqn\label{NYx}
&&\mathcal{N} (\mathbf{k},\mathbf{q})=\int \frac{d^2 \mathbf{x} }{ 2 \pi}\frac{d^2 \mathbf{y} }{ 2 \pi}  \;
e^{i \mathbf{k} \mathbf{x} } \;
e^{i (\mathbf{q}-\mathbf{k}) \mathbf{y} } \frac{\theta((\mathbf{x}-\mathbf{y})^2-c^2)}{(\mathbf{x}-\mathbf{y})^2}=
-\frac{\delta^2(\mathbf{q})}{ 2} \left(\ln \left(\frac{\mathbf{k}^2}{4}\right)-2 \psi(1) +\ln c^2\right).
\eeqn
 We have a certain freedom in choosing the cut-off $c$, which we fix by imposing a condition that the  expressions in eq.~(\ref{NYmy}) and eq.~(\ref{NYx}) are  to be identical.
This condition imposes a definite choice of the cut-off such that
 \beqn
 c^2=\frac{4 e^{2 \psi(1)-2}}{k^2}.
 \eeqn
Note that this choice of the cut-off  can be related to two basic values that govern the saturation physics. Namely,
 the coefficient that determines  the rapidity dependence of the saturation scale\footnote{For most recent review on saturation physics the reader is refrerred to the   book of Yu.~V.~Kovchegov and E.~M.~Levin and other review texts~\cite{texts}.}
 \beqn %LevinBook 4.159
Q_s(Y) \simeq Q_s(0) e^{\bar{\alpha}_s \frac{C}{2} Y}, \; C\simeq 4.88
\eeqn
and the critical exponent $\gamma_c=0.6275$   in
\beqn
N(\mathbf{x},\mathbf{y}) \propto |\mathbf{x}-\mathbf{y}|^{ 2\gamma_{c}} \; \ln  |\mathbf{x}-\mathbf{y}|,
\;\; \; \text{for} \;\;\; |\mathbf{x}-\mathbf{y}| \to 0.
\eeqn

  Those two can be expressed  at  one percent accuracy  through our choice of the cut-off  as follows
  %\beqn
  %\frac{2}{\ln^2 (c^2 k^2)} \simeq 0.6397
  %\eeqn
  \beqn
  \frac{2}{ \sqrt{c^2 k^2}} \simeq 4.84
  \eeqn
    and
\beqn
\frac{1}{\ln \left( \frac{2}{ \sqrt{c^2 k^2}}\right)} \simeq 0.634
\eeqn
establishing a relation
\beqn
\frac{1}{ \ln C} \simeq \gamma_c.
\eeqn
The authors are not  aware about any previous study discussing a similar relation.

\section{Discussions}
We discuss a new form of  the singlet BFKL eigenfunction in eq.~(\ref{Fkq}) and using it build an analytic solution of the BK equation in eq.~(\ref{BKsol}) for a particular kinematics, where the impact factor depends on the transferred momentum through eq.~(\ref{PhiNotq}). We check that the obtained solution satisfies the initial condition, it has a proper high energy behavior compatible with known values of the saturation scale and its form resembles a solution to phenomenological fan diagrams resummation.

The Balitsky-Kovchegov equation was extensively studied  during past two decades both numerically and  analytically in various approximations. The comparison  of the solution presented in this paper  to the approximate solutions available in the literature is not a simple  task  and definitely requires a separate analysis which will be published by us elsewhere.  The most important cross check  for us is the compatibility with the linear evolution of the BFKL equation and the proper high energy behaviour as shown above. We expect some peculiar technicalities
in going back to the coordinate space we have started with in eq.~(\ref{BKx}), related to the fact that we are dealing with individually divergent quantities, which nevertheless result into a finite and well defined  final answer.
The proposed analytic solution of the BK equation may have many phenomenological applications~\cite{phemo}, e.g. effective high energy  description of the proton-nucleus scattering as well as nucleus-nucleus scattering calculated through pomeron loops built of two symmetric~(projectile-target) BK solutions.

The singlet BFKL equation is reach of symmetries~\cite{lipatovdual, bondarenkoconfbk, prygarindual1, prygarindual2, sabiodual}
 and being integrable is supposed to enjoy also the dual conformal symmetry as mentioned above.
 A   similar though a more  general dual superconformal  conformal symmetry was identified for planar scattering amplitudes in maximally supersymmetric theory~(see  recent results in ref.~\cite{korchemskydual} and references wherein).
The solution of the Balitsky-Kovchegov equation presented here is written in the space of transverse momenta and one can expect  some manifistations of the dual superconformal symmetry appearing also in the  case  of the non-linear BK evolution.
We leave these issues for our  further publications.
%See  Gubser http://arxiv.org/abs/1102.4040 for more details on numerics.

\section{Acknowledgments}
The authors are indebted to I.~Balitsky, J.~Bartels, M.~Braun, E.~Gotsman,  G.~Korchemsky, E.~Levin and  L.~Lipatov for enlightening discussions.

Special thanks goes to J.~Bartels for his warm  hospitality during a stay of A.P. at University of Hamburg/DESY, where this project was initiated.
The work of A.P. on this paper was supported in part by SFB-676 Fellowship of Deutsche Forschungsgemeinschaft~(DFG).

\appendix{}
\section{ Calculation of $\mathcal{N}(\mathbf{k},\mathbf{q})|_{Y \to \infty}$}
We calculate the high energy asymptotics of the  solution in eq.~(\ref{BKsol2})
\beqn
\mathcal{N} (\mathbf{k},\mathbf{q})|_{Y \to \infty}  \simeq \sum_{n=-\infty}^{\infty} \int_{-\infty}^{\infty}  d \nu \;  \sqrt{2}  \; \chi(\nu,n)    \;F_{\nu ,n }(k,q),
\eeqn
where the function $\chi(\nu,n)$ is given in eq.~(\ref{chi}) by
\beqn
&& \chi(\nu,n) =2 \psi(1)-\psi\left(i \nu +\frac{|n|}{2}+\frac{1}{2}\right)-\psi\left(-i \nu +\frac{|n|}{2}+\frac{1}{2}\right)=
\\&&
-\sum_{m=1}^{\infty}\frac{i \nu +\frac{|n|}{2}-\frac{1}{2}}{m \left(m+i \nu +\frac{|n|}{2}-\frac{1}{2}\right)}
-\sum_{m=1}^{\infty}\frac{-i \nu +\frac{|n|}{2}-\frac{1}{2}}{m \left(m-i \nu +\frac{|n|}{2}-\frac{1}{2}\right)} \nonumber
\eeqn
and the eigenfunction $F_{\nu,n} (k,q)$  is defined in eq.~(\ref{Fkqw}) as follows
\beqn
F_{\nu,n} (k,q)=\tilde{F}_{\nu,n} (w,q)=\frac{1}{ \sqrt{2} \; \pi}\frac{|w-1|^2}{|q|^2} w^{i \nu +\frac{n}{2}-\frac{1}{2}}
{w^*}^{i \nu -\frac{n}{2}-\frac{1}{2}}.
\eeqn

Consider the following expression
\beqn
&& I=\sum_{n=-\infty}^{\infty} \int_{-\infty}^{\infty} d \nu \;  \chi(\nu, n) \; w^{i \nu +\frac{n}{2}}\; {w^*}^{i \nu -\frac{n}{2}}
=
\sum_{n=-\infty}^{\infty} \int_{-\infty}^{\infty} d \nu \;  \chi(\nu, n) \; |w|^{i 2 \nu } \; e^{i n \phi} =
\\
&&
\sum_{n=1}^{\infty} \int_{-\infty}^{\infty} d \nu \;  \chi(\nu, n) \; |w|^{i 2 \nu } \; \left(e^{i n \phi}+e^{-i n \phi}\right)
+
 \int_{-\infty}^{\infty} d \nu \;  \chi(\nu, 0) \; |w|^{i 2 \nu },
\eeqn
where we use $w/w^*=e^{i \phi}$. We do the $\nu$ integration first for $|w| >1$ closing integration  contour in the upper complex semiplane and the contribution comes only from one of the digamma functions in $\chi(\nu,n)$. Namely,
\beqn
&& I= 2 \pi \sum_{m=1}^{\infty} \sum_{n=1}^{\infty} |w|^{1-n-2 m}  \; \left(e^{i n \phi}+e^{-i n \phi}\right)
+ 2 \pi \sum_{m=1}^{\infty}  |w|^{1 -2 m}
=
2 \pi \sum_{m=1}^{\infty} |w|^{1 -2 m} \; \left(1+ \sum_{n=1}^{\infty} \left({w^*}^{-n}+{w}^{-n}\right)\right)= \nonumber
\\
&&
2 \pi \frac{|w|}{|w|^2-1}\left(\frac{1}{w^*-1}+\frac{1}{w-1}+1\right)
=2 \pi \frac{|w|}{|w-1|^2}.
\eeqn
In a similar way we close the integration contour for $\nu$ in the lower complex semiplane for $|w|<1$ and get the same result.
Using the definition of $F_{\nu,n}(k,q)$  in eq~(\ref{Fkqw}) we finally obtain

\beqn
\mathcal{N} (\mathbf{k},\mathbf{q})|_{Y \to \infty}  \simeq \sum_{n=-\infty}^{\infty} \int_{-\infty}^{\infty}  d \nu \;  \sqrt{2}  \; \chi(\nu,n)    \;F_{\nu ,n }(k,q)=\frac{|w-1|^2}{\pi |q|^2 |w|}\; I= \frac{2}{|q|^2}.
\eeqn


\begin{thebibliography}{99}



\bibitem{bfkl}
L. N. Lipatov, {\it Sov. J. Nucl. Phys.} {\bf 23}, (1976) 338;
E. A. Kuraev, L. N. Lipatov and V. S. Fadin,
{\it Sov. Phys. JETP} {\bf 45}, (1977) 199;
I. I. Balitsky and L. N. Lipatov,
{\it Sov. J. Nucl. Phys.} {\bf 28}, (1978) 822.

\bibitem{glr} L. V. Gribov, E. M. Levin and M. G. Ryskin, {\it Phys. Rep.}
{\bf 100} (1983) 1.






\bibitem{bal}
I. Balitsky, {\it Nucl. Phys.} {\bf B463} (1996) 99.


\bibitem{kov}
Y. V. Kovchegov, {\it Phys. Rev.} {\bf D60} (1999) 034008;
{\it Phys. Rev.} {\bf D61} (2000) 074018.


\bibitem{dipole}
N. N. Nikolaev and B. G. Zakharov, {\it Zeit. f\"ur. Phys.} {\bf C49} (1991)
607; {\it Phys. Lett.} {\bf B332} (1994) 184.

\bibitem{mueller}
A. H. Mueller, {\it Nucl. Phys.} {\bf B415} (1994) 373;
A. H. Mueller and B. Patel, {\it Nucl. Phys.} {\bf B425} (1994) 471;
A. H. Mueller, {\it Nucl. Phys.} {\bf B437} (1995) 107.



%\cite{Bern:2005iz}
\bibitem{bds}
  Z.~Bern, L.~J.~Dixon and V.~A.~Smirnov,
  %``Iteration of planar amplitudes in maximally supersymmetric Yang-Mills theory at three loops and beyond,''
  Phys.\ Rev.\ D {\bf 72}, 085001 (2005)
  [hep-th/0505205].
  %%CITATION = HEP-TH/0505205;%%
  %478 citations counted in INSPIRE as of 18 Mar 2015




  %\cite{Lipatov:2012gk}
\bibitem{prygarinnmhv}
  L.~Lipatov, A.~Prygarin and H.~J.~Schnitzer,
  %``The Multi-Regge limit of NMHV Amplitudes in N=4 SYM Theory,''
  JHEP {\bf 1301}, 068 (2013)
  [arXiv:1205.0186 [hep-th]].
  %%CITATION = ARXIV:1205.0186;%%
  %10 citations counted in INSPIRE as of 16 mar 2015

  %\cite{Fadin:2013sta}\bibitem{Fadin:2013sta}
  \bibitem{moebiusAdjoint}
  V.~S.~Fadin, R.~Fiore, L.~N.~Lipatov and A.~Papa,
  %``Moebius invariant BFKL equation for the adjoint representation in N=4 SUSY,''
  arXiv:1305.3395 [hep-th].
  %%CITATION = ARXIV:1305.3395;%%
  %1 citations counted in INSPIRE as of 18 mar 2015

%\cite{Fadin:2013hpa}\bibitem{Fadin:2013hpa}
  V.~S.~Fadin, R.~Fiore, L.~N.~Lipatov and A.~Papa,
  %``Mobius invariant BFKL equation for the adjoint representation in N=4 SUSY,''
  Nucl.\ Phys.\ B {\bf 874}, 230 (2013).
  %%CITATION = NUPHA,B874,230;%%
  %3 citations counted in INSPIRE as of 18 mar 2015


  %\cite{Marquet:2005zf}
\bibitem{marquet}
  C.~Marquet and G.~Soyez,
  %``The Balitsky-Kovchegov equation in full momentum space,''
  Nucl.\ Phys.\ A {\bf 760}, 208 (2005)
  [hep-ph/0504080].
  %%CITATION = HEP-PH/0504080;%%
  %36 citations counted in INSPIRE as of 16 mar 2015


























%\cite{Lipatov:1996ts}
\bibitem{texts}
  L.~N.~Lipatov,
  %``Small x physics in perturbative QCD,''
  Phys.\ Rept.\  {\bf 286}, 131 (1997)
  [hep-ph/9610276].
  %%CITATION = HEP-PH/9610276;%%
  %294 citations counted in INSPIRE as of 18 mar 2015
%\bibitem{barone}
"High-Energy Particle Diffraction", V.~Barone and  E.~Predazzi, Springer, 2002.

%\bibitem{forshaw}
"Quantum Chromodynamics and the Pomeron", J.~Forshaw and D.~Ross, Cambridge Lecture Notes in Physics, 1997.

%\bibitem{lipatovbook}
"Quantum Chromodynamics: Perturbative and Nonperturbative Aspects", B.~L.~Ioffe, V.~S.~Fadin and L.~N.~Lipatov,
Cambridge University Press, 2014.

%\bibitem{levinbook}
"Quantum Chromodynamics at High Energy", Yu.~V.~Kovchegov and E.~M.~Levin, Cambridge University Press, 2012.




\bibitem{remBDSregge}
  %\cite{Bartels:2008sc}\bibitem{sabio2}
  J.~Bartels, L.~N.~Lipatov and A.~Sabio Vera,
  %``N=4 supersymmetric Yang Mills scattering amplitudes at high energies: The Regge cut contribution,''
  Eur.\ Phys.\ J.\ C {\bf 65}, 587 (2010)
  [arXiv:0807.0894 [hep-th]].
  %%CITATION = ARXIV:0807.0894;%%
  %73 citations counted in INSPIRE as of 16 mar 2015
  %\cite{Lipatov:2010qg} \bibitem{prygarinmandel}
  L.~N.~Lipatov and A.~Prygarin,
  %``Mandelstam cuts and light-like Wilson loops in N=4 SUSY,''
  Phys.\ Rev.\ D {\bf 83}, 045020 (2011)
  [arXiv:1008.1016 [hep-th]].
  %%CITATION = ARXIV:1008.1016;%%
  %30 citations counted in INSPIRE as of 16 mar 2015

  %\cite{Lipatov:2010ad} \bibitem{prygarinsixmhv}
  L.~N.~Lipatov and A.~Prygarin,
  %``BFKL approach and six-particle MHV amplitude in N=4 super Yang-Mills,''
  Phys.\ Rev.\ D {\bf 83}, 125001 (2011)
  [arXiv:1011.2673 [hep-th]].
  %%CITATION = ARXIV:1011.2673;%%
  %33 citations counted in INSPIRE as of 16 mar 2015

 %\cite{Bartels:2010tx}\bibitem{prygarin33}
  J.~Bartels, L.~N.~Lipatov and A.~Prygarin,
  %``MHV Amplitude for 3->3 Gluon Scattering in Regge Limit,''
  Phys.\ Lett.\ B {\bf 705}, 507 (2011)
  [arXiv:1012.3178 [hep-th]].
  %%CITATION = ARXIV:1012.3178;%%
  %27 citations counted in INSPIRE as of 16 mar 2015

  %\cite{Bartels:2011nz}\bibitem{prygarinintegrable}
  J.~Bartels, L.~N.~Lipatov and A.~Prygarin,
  %``Integrable spin chains and scattering amplitudes,''
  J.\ Phys.\ A {\bf 44}, 454013 (2011)
  [arXiv:1104.0816 [hep-th]].
  %%CITATION = ARXIV:1104.0816;%%
  %18 citations counted in INSPIRE as of 16 mar 2015


  %\cite{Bartels:2011xy}\bibitem{prygarincollinear}
  J.~Bartels, L.~N.~Lipatov and A.~Prygarin,
  %``Collinear and Regge behavior of 2 -> 4 MHV amplitude in N = 4 super Yang-Mills theory,''
  arXiv:1104.4709 [hep-th].
  %%CITATION = ARXIV:1104.4709;%%
  %23 citations counted in INSPIRE as of 16 mar 2015

  %\cite{Prygarin:2011gd}\bibitem{prygarinvergu}
  A.~Prygarin, M.~Spradlin, C.~Vergu and A.~Volovich,
  %``All Two-Loop MHV Amplitudes in Multi-Regge Kinematics From Applied Symbology,''
  Phys.\ Rev.\ D {\bf 85}, 085019 (2012)
  [arXiv:1112.6365 [hep-th]].
  %%CITATION = ARXIV:1112.6365;%%
  %15 citations counted in INSPIRE as of 16 mar 2015
  %\cite{Bartels:2011ge}\bibitem{prygarin25}
  J.~Bartels, A.~Kormilitzin, L.~N.~Lipatov and A.~Prygarin,
  %``BFKL approach and $2 \to 5$ maximally helicity violating amplitude in ${\cal N}=4$ super-Yang-Mills theory,''
  Phys.\ Rev.\ D {\bf 86}, 065026 (2012)
  [arXiv:1112.6366 [hep-th]].
  %%CITATION = ARXIV:1112.6366;%%
  %18 citations counted in INSPIRE as of 16 mar 2015

  %\cite{Chachamis:2012fk}
\bibitem{Chachamis:2012fk}
  G.~Chachamis and A.~S.~Vera,
  %``The NLO N =4 SUSY BFKL Green function in the adjoint representation,''
  Phys.\ Lett.\ B {\bf 717}, 458 (2012)
  [arXiv:1206.3140 [hep-th]].
  %%CITATION = ARXIV:1206.3140;%%
  %8 citations counted in INSPIRE as of 16 mar 2015













%\cite{Mueller:1989st}
\bibitem{glaubermueller}
  A.~H.~Mueller,
  %``Small x Behavior and Parton Saturation: A QCD Model,''
  Nucl.\ Phys.\ B {\bf 335}, 115 (1990).
  %%CITATION = NUPHA,B335,115;%%
  %470 citations counted in INSPIRE as of 16 mar 2015


  \bibitem{bondarenkobound}
  S.~Bondarenko and M.~A.~Braun,
  %``Boundary conditions in the QCD nucleus-nucleus scattering problem,''
  Nucl.\ Phys.\ A {\bf 799}, 151 (2008)
  [arXiv:0708.3629 [hep-ph]].



\bibitem{rft}
%\bibitem{Bondarenko:2000uv}
  S.~Bondarenko, E.~Gotsman, E.~Levin and U.~Maor,
  %``A Pomeron approach to hadron nucleus and nucleus-nucleus t' 'soft' interactions at high-energy,''
  Nucl.\ Phys.\ A {\bf 683}, 649 (2001)
  [hep-ph/0001260].

 %\bibitem{Bondarenko:2006rh}
  S.~Bondarenko, L.~Motyka, A.~H.~Mueller, A.~I.~Shoshi and B.-W.~Xiao,
  %``On the equivalence of Reggeon field theory in zero transverse dimensions and reaction-diffusion processes,''
  Eur.\ Phys.\ J.\ C {\bf 50}, 593 (2007)
  [hep-ph/0609213].



  %\cite{Kozlov:2006cg}
\bibitem{pomeronloops}
  M.~Kozlov, E.~Levin and A.~Prygarin,
  %``The BFKL pomeron calculus: Probabilistic interpretation and high energy amplitude,''
  hep-ph/0606260.
  %%CITATION = HEP-PH/0606260;%%
  %14 citations counted in INSPIRE as of 18 mar 2015

  %\cite{Levin:2007yv}
  E.~Levin and A.~Prygarin,
  %``The BFKL Pomeron Calculus in zero transverse dimension: Summation of the Pomeron loops and the generating functional for the multiparticle production processes,''
  Eur.\ Phys.\ J.\ C {\bf 53}, 385 (2008)
  [hep-ph/0701178].
  %%CITATION = HEP-PH/0701178;%%
  %23 citations counted in INSPIRE as of 18 mar 2015

  %\cite{Kozlov:2007xc}
  M.~Kozlov, E.~Levin and A.~Prygarin,
  %``The BFKL Pomeron calculus in the dipole approach,''
  Nucl.\ Phys.\ A {\bf 792}, 122 (2007)
  [arXiv:0704.2124 [hep-ph]].
  %%CITATION = ARXIV:0704.2124;%%


  %19 citations counted in INSPIRE as of 18 mar 2015


  %\cite{Kozlov:2007xc}%\bibitem{Kozlov:2007xc}
  M.~Kozlov, E.~Levin and A.~Prygarin,
  %``The BFKL Pomeron calculus in the dipole approach,''
  Nucl.\ Phys.\ A {\bf 792}, 122 (2007)
  [arXiv:0704.2124 [hep-ph]].
  %%CITATION = ARXIV:0704.2124;%%
  %19 citations counted in INSPIRE as of 18 mar 2015

  %\cite{Levin:2007wc}\bibitem{Levin:2007wc}
  E.~Levin, J.~Miller and A.~Prygarin,
  %``Summing Pomeron loops in the dipole approach,''
  Nucl.\ Phys.\ A {\bf 806}, 245 (2008)
  [arXiv:0706.2944 [hep-ph]].
  %%CITATION = ARXIV:0706.2944;%%
  %37 citations counted in INSPIRE as of 18 mar 2015





%\bibitem{Bartels:2006ea}
\bibitem{phemo}
  J.~Bartels, S.~Bondarenko, K.~Kutak and L.~Motyka,
  %``Exclusive Higgs boson production at the LHC: Hard rescattering corrections,''
  Phys.\ Rev.\ D {\bf 73}, 093004 (2006)
  [hep-ph/0601128].
%\bibitem{Bondarenko:2006ft}
  S.~Bondarenko and L.~Motyka,
  %``Solving effective field theory of interacting QCD pomerons in the semi-classical approximation,''
  Phys.\ Rev.\ D {\bf 75}, 114015 (2007)
  [hep-ph/0605185].	

































%\cite{Lipatov:1998as}
\bibitem{lipatovdual}
  L.~N.~Lipatov,
  %``Duality symmetry of Reggeon interactions in multicolor QCD,''
  Nucl.\ Phys.\ B {\bf 548}, 328 (1999)
  [hep-ph/9812336].
  %%CITATION = HEP-PH/9812336;%%
  %61 citations counted in INSPIRE as of 16 mar 201





%\cite{Prygarin:2009tn}
\bibitem{prygarindual1}
  A.~Prygarin,
  %``Duality symmetry in high energy scattering,''
  arXiv:0908.2386 [hep-ph].
  %%CITATION = ARXIV:0908.2386;%%
  %1 citations counted in INSPIRE as of 16 mar 2015

%\cite{Prygarin:2009zz}
\bibitem{prygarindual2}
  A.~Prygarin,
  %``Duality symmetry of BFKL equation: Reggeized gluons versus color dipoles,''
  Phys.\ Rev.\ C {\bf 83}, 055206 (2011)
  [arXiv:0911.5279 [hep-ph]].
  %%CITATION = ARXIV:0911.5279;%%
  %1 citations counted in INSPIRE as of 16 mar 2015



%\cite{Gomez:2009bx}
\bibitem{sabiodual}
  C.~Gomez, J.~Gunnesson and A.~S.~Vera,
  %``Dual conformal invariance in the Regge limit,''
  Phys.\ Lett.\ B {\bf 690}, 78 (2010)
  [arXiv:0908.2568 [hep-th]].
  %%CITATION = ARXIV:0908.2568;%%
  %3 citations counted in INSPIRE as of 16 mar 2015

%\cite{Bondarenko:2007sm}
\bibitem{bondarenkoconfbk}
  S.~Bondarenko and A.~Prygarin,
  %``BFKL ansatz for BK equation in conformal basis,''
  Nucl.\ Phys.\ A {\bf 800}, 63 (2008)
  [arXiv:0709.3010 [hep-ph]].
  %%CITATION = ARXIV:0709.3010;%%
  %4 citations counted in INSPIRE as of 16 mar 2015












%\cite{Derkachov:2013bda}
\bibitem{korchemskydual}
  S.~É.~Derkachov, G.~P.~Korchemsky and A.~N.~Manashov,
  %``Dual conformal symmetry on the light-cone,''
  Nucl.\ Phys.\ B {\bf 886}, 1102 (2014)
  [arXiv:1306.5951 [hep-th]].
  %%CITATION = ARXIV:1306.5951;%%
  %3 citations counted in INSPIRE as of 16 mar 2015
















































\end{thebibliography}
\end{document}